\documentclass[prd,aps,nofootinbib,singlecolumn]{revtex4-2}
\usepackage{amsmath,amssymb,amsfonts,color,graphicx,graphics,latexsym,placeins,epsfig,multirow}
\usepackage[toc]{appendix}
\usepackage{array}
\newcolumntype{P}[1]{>{\centering\arraybackslash}p{#1}}
\usepackage{footmisc}
\usepackage{mathrsfs}
\usepackage[compatibility=false]{caption}
\usepackage{subcaption}
\usepackage{comment}
\usepackage{bbold}
\usepackage{enumitem}
\setdescription{leftmargin=12.5pt}
\usepackage{bbold}
\usepackage{hyperref}
\usepackage{varioref}
\usepackage{color}
\def\a{\alpha}

\def\6{\partial}

\begin{document}

\title{\Large \bf Action principle of Galilean relativistic Proca theory}
\author{Rabin Banerjee \footnote{DAE Raja Ramanna fellow}}
\author{Soumya Bhattacharya}
\affiliation{Department of Astrophysics and High Energy Physics, S.N. Bose National Center for Basic Sciences, Kolkata 700106, India}
\email{rabin@bose.res.in}
\email{soumya557@bose.res.in}


\vskip .2in
\begin{abstract}
     In this paper, we discuss Galilean relativistic Proca theory in detail. We first provide a set of mapping relations, derived systematically, that connect the covariant and contravariant vectors in the Lorentz relativistic and Galilean relativistic formulations. Exploiting this map, we construct the two limits of Galilean relativistic Proca theory from usual Proca theory in the potential formalism for both contravariant and covariant vectors which are now distinct entities. An action formalism is thereby derived from which the  field equations are obtained and their internal consistency is shown.  Next we construct Noether currents and show their on-shell conservation. We introduce analogues of Maxwell's electric and magnetic fields and recast the entire analysis in terms of these fields. Explicit invariance under Galilean transformations is shown for both electric/magnetic limits. We then move to discuss Stuckelberg embedded Proca model in the Galilean framework.

\end{abstract}
\maketitle
\section{Introduction}
\noindent Works on non-Lorentzian physics have gained considerable attention recently. Non-relativistic limit or Galilean relativistic limit of various field theories and gauge theories have been studied extensively.  These non-relativistic theories have found applications in a wide variety of fields including but not restricted to holography \cite{Taylor}, non-relativistic diffeomorphisms (NRDI) \cite{andreev1, andreev2, jensen, rb1, rb2}, 
condensed matter systems \cite{Pal}, fluid dynamics \cite{Jain, rb3}, gravitational waves \cite{Morand}. This formulation is tricky and quite different from the usual relativistic
case. Covariance in non-relativistic physics is subtle due to the absolute nature of time. The lack of a single non-degenerate metric in the non-relativistic case poses some additional difficulties. The study of Galilean relativistic theories was initiated by Le Bellac and Levy Leblond \cite{Leblond} back in 1970's.\footnote{Also, see ref \cite{Sengupta}.} Their main focus was on Galilean electromagnetism which had many physical applications \cite{Germain}. The experiments of Rowland, Vasilescu, Karpen, Roentgen, Eichenwald and others, reviewed in \cite{Germain}, which were previously interpreted or understood using special relativity found a more eloquent explanation using Galilean electromagnetism since it was perfectly suited to study electrodynamics of continuous media at low velocities. It may also be recalled that the results in \cite{Leblond} were attained primarily by physical principles based on gauge and galilean invariances. In case of non-gauge theories such an approach would be untenable, or at least difficult to formulate.\\
\indent An alternative approach would be to construct the Galilean relativistic version by taking an appropriate limit of the corresponding relativistic theory. Such attempts were done in \cite{Duval}, using group contraction techniques. Likewise in \cite{Mehra1}, scaling relations among potentials were introduced to obtain equations of motion in Galilean electrodynamics from Maxwell electrodynamics. Recently we \cite{Bhattacharya} have developed a method where the Galilean relativistic version of any Lorentz relativistic vector theory can be constructed following a structured algorithm. It does not depend whether the vector theory is a gauge or non-gauge theory. The method was applied to the Maxwell theory \cite{Bhattacharya}.\\
\indent In this paper, we are interested in Proca theory which describes a massive spin-1 field in Minkowski spacetime \cite{Proca}. The presence of the mass term actually breaks gauge invariance of the lagrangian. Proca theory plays a key role in many areas of fundamental physics, including experimental physics. For example it plays an important role in the theoretical framework for experimental studies aiming at the determination of upper bounds for the mass of the photon \cite{Luo}, \cite{Nieto}, and other different areas of fundamental physics \cite{Sampaio, Dvali, Tomaschitz}. Apart from these there is a recent resurgence of the study of Proca model from different aspects. For example there are studies in the context of gravitational waves \cite{Radu1}, gravitational lensing and black hole (BH) shadows \cite{Radu2}, alternative theories of gravity \cite{Demir}. There has been a prescription to generalise the Proca action \cite{Heisenberg}. This generalised Proca model has various applications in different areas of fundamental physics \cite{Said, Pichet}. There is also a non-linear Proca model called Proca-Nuevo model \cite{derham}. So given this importance and applications of Proca model we believe the study of non-relativistic limits of various Proca models will be interesting. Our current work is the first step towards that direction.  \\
\indent The construction of the cherished action and the associated action principle follows by explicitly providing maps that connect the relativistic with non-relativistic vectors. This is done for both covariant and contravariant vectors since these are distinct entities in the Galilean theory, not being connected by any metric. Also for each component, there are maps corresponding to electric and magnetic limits - the two limits in Galilean relativistic physics. Using these maps the lagrangian for Galilean relativistic Proca model is constructed from its relativistic counterpart, both for electric and magnetic limits, and its consequences are examined. \\
\indent The paper is organised as follows: in section \ref{sec2} we derive mapping relations between relativistic and non-relativistic vectors for electric and magnetic limit for both contravariant and covariant vectors. In section \ref{sec3} we derive the non-relativistic lagrangian for both limits and write down the equations of motion from an action principle. These are shown to be compatible with the equations derived directly from the relativistic Proca equations. In section \ref{sec4} we introduce the galilean electric and magnetic fields and write down the Galilean Proca equations. The Noether currents and their corresponding conservations are discussed in section \ref{sec5}. We now move to discuss the gauge invariant version of Proca theory, obtained by Stuckelberg embedding, and perform its detailed Galilean analysis. This has been discussed in section \ref{sec6}. Finally we conclude in section \ref{sec7}.
\section{Maps relating Lorentz and Galilean vectors} \label{sec2}
\noindent Here we derive a certain scaling between special relativistic and Galilean relativistic quantities. As we know there exists two types of such limits for the vector quantities namely electric and magnetic limits. So first let us consider the contravariant vectors. Let us take a  generic Lorentz transformation with the boost velocity as $u^i$:
\begin{equation}
    x'^0 = \gamma x^0 - \frac{\gamma u_i}{c} x^i
    \label{t1}
\end{equation}
\begin{equation}
    x'^i = x^i - \frac{\gamma u^i}{c}x^0 + (\gamma -1 )\frac{u^i u_j}{u^2}x^j
    \label{t2}
\end{equation}
where $\gamma = \frac{1}{\sqrt{1-\frac{u^2}{c^2}}}$. Under such Lorentz transformations a contravariant vector changes as
\begin{equation*}
    V'^{\mu} = \frac{\6 x'^{\mu}}{\6 x^{\nu}} V^{\nu}
\end{equation*}
We can write them component-wise as (also considering $u<<c$, so $\gamma \to 1$)
\begin{equation}
    V'^0 = V^0 - \frac{u_j}{c}V^j
    \label{contra1}
\end{equation}
\begin{equation}
    V'^i = V^i - \frac{u^i}{c} V^0
    \label{contra2}
\end{equation}
We next provide a map that relates the Lorentz vectors with their Galilean counterparts. 
\footnote{Notation: Here relativistic vectors are denoted by capital letters ($V^0, V^i$ etc) and Galilean vectors are denoted by lowercase letters ($v^0, v^i$ etc).} 
\begin{equation}
    V^0 = c v^0, \,\,\,\, V^i = v^i
    \label{contrael}
\end{equation}
This particular map corresponds to the case $\frac{V^0}{V^i} = c ~\frac{v^0}{v^i}$ in the $c \to \infty$ limit. This yields largely timelike vectors and is called 'electric limit'.
Now using eqn \ref{contrael} in eqns \ref{contra1} and \ref{contra2} we get
\begin{equation}
v'^0 = v^0
\label{v1}
\end{equation}
\begin{equation}
    v'^i = v^i - u^i v^0
    \label{v2}
\end{equation}
These equations define the usual galilean transformations. To see this in the context of coordinates, we revert to (\ref{t1}, \ref{t2}), consider the $u^2 << c^2$ limit implying $\gamma \to 1$, use $x^0 = ct$ so that, 
\begin{equation}
    t' = t, \,\,\, x'^i = x^i - u^i t \label{a1}
\end{equation}
which is the exact analogue of (\ref{v1}, \ref{v2}).\\
\noindent We next consider the magnetic limit which corresponds to largely spacelike vectors
\begin{equation}
    V^0 = -\frac{v^0}{c}, \,\,\,\, V^i = v^i
    \label{contramag}
\end{equation}
Now using \ref{contramag} in \ref{contra1} and \ref{contra2} we get

\begin{equation}
    v'^0 = v^0 + u_j v^j
    \label{v3}
\end{equation}
\begin{equation}
    v'^i = v^i
    \label{v4}
\end{equation}
which is again a galilean transformation, although the corresponding group is not an invariance group of classical mechanics. Expectedly, the role of time-like and space-like vectors has been reversed from the earlier case. \\ 
\indent We will now consider the covariant vectors. We will write first the reverse transformations of eqn \ref{t1} and \ref{t2} which are
\begin{equation}
    x^0 = \gamma x'^0 + \frac{\gamma u_i}{c} x'^i
    \label{t3}
\end{equation}
\begin{equation}
    x^i = x'^i + \frac{\gamma u^i}{c}x'^0 + (\gamma -1 )\frac{u^i u_j}{u^2}x'^j
    \label{t4}
\end{equation}
Covariant vectors transform as
\begin{equation*}
    V'_{\mu} = \frac{\6 x^{\nu}}{\6 x'^{\mu}} V_{\nu}
\end{equation*}
Componentwise we write them as
\begin{equation}
    V'_0 = V_0 + \frac{u^i}{c} V_i
    \label{cov1}
\end{equation}
\begin{equation}
    V'_i = V_i + \frac{u_i}{c} V_0
    \label{cov2}
\end{equation}
Now here we take the  electric limit in the following way, which will soon become clear
\begin{equation}
    V_0 = \frac{v_0}{c}, \,\,\,\, V_i = v_i
    \label{covel}
\end{equation}
Using \ref{covel} in \ref{cov1} and \ref{cov2} we get
\begin{eqnarray}
  v'_0 = v_0 + u^i v_i
  \label{v5}
\end{eqnarray}
\begin{eqnarray}
  v'_i = v_i
  \label{v6}
\end{eqnarray}
which are again Galilean transformations.
We will now consider the magnetic limit as 
\begin{equation}
    V_0 = -c v_0, \,\,\,\, V_i = v_i
    \label{covmag}
\end{equation}
Using \ref{covmag} in \ref{cov1} and \ref{cov2} we get 
\begin{equation}
    v_0' = v_0
    \label{v7}
\end{equation}
\begin{equation}
    v'_i = v_i - u_i v_0 
    \label{v8}
\end{equation}
 The mapping relations are summarised in the table \ref{T1}.
\begin{table}
\caption{Mapping relations}\label{T1}
\begin{center}
\begin{tabular}{|c|c|c|} \hline 
${\rm Limit}$  & $ {\rm Contravariant~~mapping}$ & $ {\rm Covariant~~mapping}$  \\ \hline
${\rm Electric ~~limit}$ & $V^0 \to c~v^0, \,\,  V^i \to v^i$ & $V_0 \to \frac{v_0}{c}, \,\, V_i \to v_i$ \\ \hline
${\rm Magnetic ~~limit}$ & $V^0 \to -\frac{v^0}{c},\,\,V^i \to v^i $ & $V_0 \to -c~v_0. \,\, V_i \to v_i$ \\
\hline
\end{tabular}
\label{T1}
\end{center}
\end{table}

\section{Lagrangian and field equations} \label{sec3}
\noindent Now let us start from the relativistic Proca theory described by the Lagrangian
\begin{equation}
    \mathcal{L} = -\frac{1}{4}~F_{\mu \nu}F^{\mu \nu} + \frac{k^2}{2} A_\mu A^\mu
    \label{l0}
\end{equation}
where $F_{\mu \nu} = \partial_{\mu} A_{\nu} - \partial_{\nu} A_{\mu}$ and $\eta_{\mu \nu}$ is the flat space metric with signature $\Big(-,+,+,+\Big)$. 

It gives rise to the following eqns of motion
\begin{eqnarray}
    \6^\mu F_{\mu \nu} + k^2 A_\nu = 0 \label{peom1}
\end{eqnarray}
which imply
\begin{eqnarray}
    \6^{\nu} \Big( \6^\mu F_{\mu \nu} + k^2 A_\nu \Big) = 0 \implies \6^\mu A_\mu = 0 
    \label{peom2}
\end{eqnarray}
Eqn \ref{peom2} is a necessary condition for the relativistic Proca theory, leading to a Klein-Gordon type equation,
\begin{equation}
  \Big(\Box + k^2  \Big) A_\nu = 0  \label{rwe}
\end{equation}
\subsection{Galilean relativistic theory}
We now discuss the derivation of the Galilean invariant Proca model using results of section 2. The two limits are now considered independently. The first step is to open the various terms in \ref{l0} as,
\begin{equation}
    \mathcal{L} = -\frac{1}{4}\Big(2F_{0i}F^{0i} + F_{ij}F^{ij} \Big) + \frac{k^2}{2} \Big(A_0 A^0 + A_i A^i  \Big)
    \label{l1}
\end{equation}

\noindent {\underline {\bf Electric limit:}}\\
Using the relations given in table \ref{T1} we  write the two terms in \ref{l1} as, 
\begin{eqnarray}
2 F_{0i}F^{0i}  \xrightarrow[\text{$c \to \infty$}]{\text{electric~limit}}   -2 \6^i a^0 \Big( \partial_t a_i - \partial_i a_0  \big) \hspace{1in}
\label{part1}
\end{eqnarray}
 
\begin{eqnarray}
F_{ij}F^{ij}  \xrightarrow[\text{$c \to \infty$}]{\text{electric~limit}} \Big(\partial_i a_j - \partial_j a_i \Big) \Big(\partial^i a^j - \partial^j a^i \Big) \equiv f_{ij} f^{ij} 
\end{eqnarray}

\begin{eqnarray}
\frac{k^2}{2} \Big(A_0 A^0 + A_i A^i  \Big)  \xrightarrow[\text{limit}]{\text{electric}}   \frac{k^2}{2} \Big(a_0 a^0 + a_i a^i  \Big) 
\end{eqnarray}

\noindent Here $A^{\mu}$ is the relativistic four potential while $a^0$ and $a^i$ are it's galilean counterpart. Hence the full lagrangian takes the following form
\begin{equation}
    \mathcal{L}_e = \frac{1}{2} \6^i a^0 \Big( \partial_t a_i - \partial_i a_0  \big)  - \frac{1}{4} f_{ij} f^{ij} + \frac{k^2}{2} \Big(a_0 a^0 + a_i a^i  \Big) 
    \label{el1}
\end{equation}
and the corresponding action is,
\begin{equation}
    S_e = \int d^3 x dt \mathcal{L}_e
\end{equation}
\noindent Now we derive the equations of motion. Variation with respect to $a_0,~a_j,~a^0,~a^j$ yields the equations of motion
\begin{eqnarray}
\6_i \6^i a^0 + k^2 a^0 = 0 
\label{ee1}
\end{eqnarray}
\begin{eqnarray}
\6_t \6^j a^0 + \6_i \6^j a^i - \6_i \6^i a^j - k^2 a^j = 0
\label{ee2}
\end{eqnarray}
\begin{eqnarray}
\6^i \6_t a_i - \6^i \6_i a_0 - k^2 a_0 = 0
\label{ee3}
\end{eqnarray}
\begin{eqnarray}
\6^i \6_i a_j - \6^i \6_j a_i + k^2 a_j = 0 
\label{ee4}
\end{eqnarray}
We now derive these equations directly from the relativistic equations of motion \ref{peom1}, which can be split as
\begin{eqnarray}
\6_i F^{i0} + k^2 A^0 =0 \label{e1}\\
\6_0 F^{0j} + \6_i F^{ij} + k^2 A^j = 0 \label{e2}
\end{eqnarray}
From table \ref{T1} and eqn \ref{e1} we get in the electric limit
\begin{eqnarray}
\6_i F^{i0} + k^2 A^0  =0  
\xrightarrow[\text{$c \to \infty$}]{\text{electric~limit}} \6_i \6^i a^0 + k^2 a^0 = 0 
\label{eo1}
\end{eqnarray}
which reproduces eqn \ref{ee1}. Likewise the Galilean limit of \ref{e2} reproduces \ref{ee2}.

To get the remaining pair of equations \ref{ee3} and \ref{ee4}  we have to interpret  eqns \ref{e1} and \ref{e2} in terms of their covariant components, 
\begin{eqnarray}
\6^i F_{i0} + k^2 A_0 =0 \label{e3}\\
\6^0 F_{0j} + \6^i F_{ij} +k^2 A_j = 0 \label{e4}
\end{eqnarray}
Once again the Galilean version is obtained from table \ref{T1}, followed by taking $c \to \infty$. Equations \ref{ee3} and \ref{ee4} are reproduced.
This shows the consistency of the eqn of motion in Galilean Proca model.\\
\indent Let us now show how \ref{peom2} is manifested in the present analysis. Since covariant and contravariant components are treated separately in the Galilean construction, \ref{peom2} is interpreted as, 
\begin{equation}
    \6^\mu A_\mu = \6_\mu A^\mu = 0
\end{equation}
These conditions imply, 
\begin{eqnarray}
    \6^\mu A_\mu = 0 \implies 
    -\frac{1}{c} \6_t \Big(\frac{1}{c} a_0 \Big) + \6^i a_i = 0 \xrightarrow[\text{$c \to \infty$}]{\text{}} \6^i a_i = 0 \label{ae1}
\end{eqnarray}
and,
\begin{eqnarray}
    \6_\mu A^\mu = 0 \implies \6_t a^0 + \6_i a^i = 0 \label{ae2}
\end{eqnarray}
The relation \ref{ae1} follows on acting $\6^j$ on either side of \ref{ee4}. Likewise \ref{ae2} follows on acting $\6_j$ on \ref{ee2} and using \ref{ee1}. We will now discuss the implications of eqn \ref{rwe}. To do that we can write for its  contravariant components, 
\begin{equation}
    \Big(-\frac{1}{c^2} \6_t^2 + \6_j \6^j + k^2 \Big) A^\nu =0 \label{bstr}
    \end{equation}
Considering the electric limits for $\nu = 0 ~\& ~i$ we have following two equations
\begin{eqnarray}
    \6_i \6^i a^0 + k^2 a^0 = 0, \label{ewe1} \\
    \6_i \6^i a^j + k^2 a^j = 0 \label{ewe2}
\end{eqnarray}
Now eqn \ref{ewe1} is nothing but \ref{ee1}. We can re-write eqn \ref{ee2} as
\begin{equation}
    \6^j \Big(\6_t a^0 + \6_i a^i  \Big) - \6_i \6^i a^j - k^2 a^j = 0 \implies \Big(\6_i \6^i + k^2 \Big) a^j = 0 
\end{equation}
which reproduces eqn \ref{ewe2}. The term in the bracket vanishes as a consequence of \ref{ae2}. We can also write the other two equations i.e \ref{ee3} and \ref{ee4} as the the form given by \ref{rwe}. \\
\noindent {\underline {\bf Magnetic limit:}}\\
Here again using the relations given in table \ref{T1} we can write the two terms in \ref{l1} as,
\begin{eqnarray}
2 F_{0i}F^{0i}  \xrightarrow[\text{$c \to \infty$}]{\text{magnetic limit}} -2 \6_i a_0 \Big(\6_t a^i - \6^i a^0 \Big) \hspace{1in}
\end{eqnarray}
\begin{eqnarray}
F_{ij}F^{ij} 
\xrightarrow[\text{$c \to \infty$}]{\text{magnetic limit}} f_{ij}f^{ij} \hspace{1in} 
\end{eqnarray}
\begin{eqnarray}
\frac{k^2}{2} \Big(A_0 A^0 + A_i A^i  \Big)  \xrightarrow[\text{$c \to \infty$}]{\text{magnetic~limit}}   \frac{k^2}{2} \Big(a_0 a^0 + a_i a^i  \Big) 
\end{eqnarray}
So the magnetic action will take the following form
\begin{equation}
    S_m = \int d^3 x dt \mathcal{L}_m
\end{equation}
where, 
\begin{equation}
\mathcal{L}_m = \frac{1}{2}  \6_i a_0 \Big(\6_t a^i - \6^i a^0 \Big) - \frac{1}{4} f_{ij}f^{ij} + \frac{k^2}{2} \Big(a_0 a^0 + a_i a^i  \Big)  \label{l2}
\end{equation}
Variation  wrt $a_0,~a_j,~a^0,~a^j$ we get,
\begin{eqnarray}
   \partial_i ~\partial_t~ a^i - \partial^i ~\partial_i ~a^0 - k^2 a^0 = 0 
  \label{em1}
\end{eqnarray}
\begin{eqnarray}
   \6_i \6^i a^j - \6_i \6^j a^i + k^2 a^j = 0
  \label{em2}
  \end{eqnarray}
\begin{eqnarray}
  \6^i \6_i a_0 + k^2 a_0 = 0
  \label{em3}
\end{eqnarray}
\begin{eqnarray}
 \6_t \6_j a_0 + \6^i \6_j a_i - \6^i \6_i a_j - k^2 a_j = 0 \label{em4}
\end{eqnarray}
Here also we can show that the above equations agree with those derived directly from relativistic Proca equations by taking the magnetic limit.
From \ref{e1} we have
\begin{eqnarray}
\6_i \Big( -\frac{1}{c}\6^i a^0 + \frac{1}{c} \6_t a^i \Big) - \frac{1}{c} k^2 a^0 = 0 \nonumber  \xrightarrow[\text{$c \to \infty$}]{\text{magnetic limit}} \6_i \6_t a^i - \6_i \6^i a^0 - k^2 a^0 = 0
\label{eo5}
\end{eqnarray}
which reproduces eqn \ref{em1}. 
Similarly \ref{e2} reproduces \ref{em2}
\\
\noindent To get the remaining pair of equations we have to start from the covariant versions (\ref{e3}, \ref{e4}) which will reproduce \ref{em3} and \ref{em4} respectively. 

\begin{table}
\caption{Field equations}\label{T2}
\begin{center}
\begin{tabular}{|c|c|c|} \hline 
${\rm Variables}$ & ${\rm Electric ~~limit}$ & ${\rm Magnetic ~~limit}$ \\ \hline
$a^0$  & $\6^i \6_t a_i - \6^i \6_i a_0 -k^2 a_0 = 0$ & $\6^i \6_i a_0 +k^2 a_0 = 0$ \\ \hline
$a^j$ & $\6^i \6_i a_j - \6^i \6_j a_i + k^2 a_j = 0  $ & $\6_t \6_j a_0 + \6^i \6_j a_i - \6^i \6_i a_j - k^2 a_j = 0$ \\ \hline
$a_0$ & $ \partial^i \partial_i a^0 +k^2 a^0 = 0 $ & $\partial_i ~\partial_t~ a^i - \partial^i ~\partial_i ~a^0 - k^2 a^0 = 0 $ \\ \hline
$a_j$ & $\6_t \6^j a^0 + \6_i \6^j a^i - \6_i \6^i a^j - k^2 a^j = 0$ & 
$ \6_i \6^i a^j - \6_i \6^j a^i + k^2 a^j = 0 $ \\ \hline
\end{tabular}
\label{T2}
\end{center}
\end{table}
\noindent The field equations for both limits of Galilean Proca model are shown in table \ref{T2}.\\


\noindent In magnetic limit the condition \ref{peom2} implies
\begin{eqnarray}
 \6^\mu A_\mu = 0 
\implies -\frac{1}{c} \6_t \Big( -c a_0 \Big) + \6^i a_i =0 \xrightarrow[\text{$c \to \infty$}]{\text{}} \6_t a_0 + \6^i a_i = 0 \label{am1}
\end{eqnarray}
Similarly,
\begin{eqnarray}
    \6_\mu A^\mu = 0 \implies \6_i a^i = 0 \label{am2}
\end{eqnarray}
Here also the relation \ref{am1} follows on acting $\6^j$ on eqn \ref{em4} and using eqn \ref{em3}. Similarly relation \ref{am2} follows on acting $\6_j$ on \ref{em2}.  We will now discuss the implications of eqn \ref{rwe} in the magnetic limit. For the contravariant components the equations are
\begin{eqnarray}
    \Big(\6_i^2 + k^2  \Big) a^0 = 0 \label{mew1}\\
    \Big(\6_i^2 + k^2   \Big) a^i =0 \label{mew2}
\end{eqnarray}
From the equations of motion, \ref{em1} implies
\begin{equation}
    \6_t \Big(\6_i a^i \Big) - \6_i \6^i a^0 - k^2 a^0 = 0 \implies \Big( \6_i^2 + k^2 \Big) a^0 =0 \label{mw1}
\end{equation}
In the first line $\6_i a^i = 0 $ from \ref{am2}. Eqn \ref{mw1} reproduces \ref{mew1}. Similarly \ref{em2} reproduces \ref{mew2} by exploiting \ref{am2}. The other two equation \ref{em3} and \ref{em4} can be written in the form of \ref{rwe} by exploiting \ref{am1}. \\
\indent From table \ref{T2} we observe the dual role of electric and magnetic limits vis-a-vis their covariant and contravariant sectors. In other words the electric limit equation found by varying $a^0$ corresponds to the magnetic limit equation obtained by varying $a_0$. The same property holds for other variables. 
\section{Electric and magnetic fields in Galilean Proca theory} \label{sec4}
\noindent Here we discuss Galilean Proca theory in terms of electric and magnetic fields introduced in analogy with Maxwell theory. For this purpose we will discuss contravariant and covariant sectors separately.
\subsection{Contravariant sector}
Relativistic electric and magnetic fields are defined as 
\begin{eqnarray}
E^i = \6^0 A^i - \6^i A^0 \\
B^i = \epsilon^{ij}_{~k} \6_j A^k
\end{eqnarray}
First, we consider the electric limit. \\
\noindent \underline {\bf Electric limit:}\\
\noindent Using the mapping relations given in table \ref{T1} we can write the electric field as 
\begin{eqnarray}
E^i = -\frac{1}{c} \6_t a^i - c \6^i a^0 
\end{eqnarray}
We can define the Galilean electric and magnetic fields as 
\begin{equation}
    e^i = \lim_{c \to \infty} \frac{E^i}{c} = -\6^i a^0, \,\,\,~~~~~ b^i =  \lim_{c \to \infty} B^i = \epsilon^{ij}_{~k} \6_j a^k
\end{equation}

\noindent    Now we write the field equations that we derived in the previous section  in terms of these electric and magnetic fields. From \ref{ee1} we get 
    \begin{eqnarray}
    \6_i \6^i a^0 = -k^2 a^0 \implies \6_i (-e^i) = -k^2 a^0 \implies \vec \nabla . \vec e = k^2 a^0
    \end{eqnarray}
\noindent    Similarly eqn \ref{ee2} implies
    \begin{eqnarray}
   \6_t \6^j a^0 + \6_i \6^j a^i - \6_i \6^i a^j -k^2 a^j= 0 
     \implies  (\vec \nabla \times \vec b)^j = \6_t e^j + k^2 a^j
    \end{eqnarray}
    
\noindent We can see clearly that
\begin{equation}
 \vec \nabla .\vec b =  \6_i b^i = \6_i \epsilon^{ij}_{~k} \6_j a^k =\epsilon^{ij}_{~k} \6_i\6_j a^k = 0  
\end{equation}

\noindent We will now compute $\vec \nabla \times \vec e$,
\begin{eqnarray}
(\nabla \times e)^i = \epsilon^{ij}_{~k} \6_j e^k 
= \epsilon^{ij}_{~k} \6_j (-\6^k a^0) = 0 
\end{eqnarray}
So in electric limit we get the following set of equations
\begin{eqnarray}
\vec \nabla . \vec e = \6_i e^i = k^2 a^0, \label{max01}\\
\vec \nabla . \vec b =\6_i b^i = 0, \label{max02}\\ 
(\vec \nabla \times \vec e)^i = \epsilon^{ij}_{~k} \6_j e^k = 0 \label{max1} \\
(\vec \nabla \times \vec b)^i = \epsilon^{ij}_{~k} \6_j b^k = \6_t (\vec e)^i + k^2 a^i\label{max2}
\end{eqnarray}

\noindent \underline {\bf Magnetic limit:}\\
\noindent Electric field can be written in this limit from the mapping relations in table \ref{T1} as 
\begin{equation}
    E^i = -\frac{1}{c} \6_t a^i + \frac{1}{c} \6^i a^0
\end{equation}
We define the galilean electric and magnetic fields as 
\begin{equation}
    e^i = \lim_{c \to \infty} c E^i = -(\6_t a^i - \6^i a^0), \,\,\, ~~~~~ b^i = \lim_{c \to \infty} B^i = \epsilon^{ij}_{~k} \6_j a^k
\end{equation}
Using these relations, eqns \ref{em1}, \ref{em2} and two Bianchi identities are expressed in terms of electric/magnetic fields as

\begin{eqnarray}
\vec \nabla . \vec e = \6_i e^i = -k^2 a^0, \label{max03} \\
\vec \nabla . \vec b =\6_i b^i = 0, \label{max04}\\ 
(\vec \nabla \times \vec e)^i = \epsilon^{ij}_{~k} \6_j e^k = -\6_t (\vec b)^i, \label{max3} \\
(\vec \nabla \times \vec b)^i = \epsilon^{ij}_{~k} \6_j b^k  = k^2 a^i \label{max4}
\end{eqnarray}
\noindent The results for the electric/magnetic fields as well as the Maxwell-type equations (\ref{max01}, \ref{max02}, \ref{max1}, \ref{max2}) and (\ref{max03}, \ref{max04}, \ref{max3}, \ref{max4}) for either electric or magnetic limit agree with those given in \cite{Santos}. In \cite{Santos} an additional mass term has been introduced to express the Galilean algebra as a centrally extended Bargmann algebra. Correspondingly, no superselection rule is needed and one works with an ordinary representation. In our analysis we are confined to the Galilean algebra but the ordinary representation has to be replaced by a projective representation. The two descriptions are completely equivalent \cite{Weinberg}. 
\subsection{Covariant sector}
Relativistic electric and the magnetic fields are defined as 
\begin{eqnarray}
E_i = - \Big( \6_0 A_i - \6_i A_0\Big) \\
B_i = \epsilon_i^{~jk} \6_j a_k
\end{eqnarray}
First, we consider the electric limit. \\
\noindent \underline {\bf Electric limit:}\\
In this limit the electric field looks like 
\begin{equation}
E_i = -\Big(\frac{1}{c} \6_t a_i + \frac{1}{c} \6_i   a_0\Big)
\end{equation}
The Galilean electric and magnetic fields are given by,  
\begin{equation}
    e_i = \lim_{c \to \infty} c E_i = - (\6_t a_i - \6_i a_0), \,\,\,~~~~~~ b_i = \lim_{c \to \infty} B_i = \epsilon_{i}^{~jk} \6_j a_k
\end{equation}
Using these expressions the Proca equations in the Galilean electric limit are found to be, 
\begin{eqnarray}
\vec \nabla . \vec e = \6^i e_i = -k^2 a_0 \\
\vec \nabla . \vec b = \6^i b_i = 0\\ 
\Big(\vec \nabla \times \vec e\Big)_i = \epsilon_{i}^{~jk}\6_j e_k = - \6_t (\vec b)_i \\
\Big(\vec \nabla \times \vec b \Big)_i= \epsilon_{ij}^{~~k} \6^j b_k = k^2 a_i
\end{eqnarray}
\noindent \underline {\bf Magnetic limit:}\\
In this limit electric field is scaled as 
\begin{equation}
    E_i = -\Big(\frac{1}{c} \6_t a_i + c \6_i a_0 \Big)
\end{equation}
The Galilean electric and magnetic fields are now defined as,
\begin{equation}
    e_i = \lim_{c \to \infty} \frac{E_i}{c} = -\6_i a_0, \,\,\, ~~~~~ b_i = \lim_{c \to \infty} B_i = \epsilon_{i}^{~jk} \6_j a_k
\end{equation}
 Finally, the equations we get in the magnetic limit are given by,
\begin{eqnarray}
\vec \nabla . \vec e = \6^i e_i = k^2 a_0 \\
\vec \nabla . \vec b = \6^i b_i = 0\\ 
\Big(\vec \nabla \times \vec e \Big)_i = \epsilon_{i}^{~jk} \6_j e_k = 0 \\
\Big(\vec \nabla \times \vec b \Big)_i = \epsilon_{ij}^{~~k} \6^j b_k = \6_t e_i + k^2 a_i
\end{eqnarray}
\subsection{Lagrangian in field formulation and boost invariance}
\noindent In both electric and magnetic limits the lagrangian will take the following form 
\begin{equation}
    \mathcal{L}_e = \mathcal{L}_m =  \frac{1}{2} \Big(e^i e_i - b_i b^i \Big) + \frac{k^2}{2} \Big(a_0 a^0 + a_i a^i  \Big)  \label{lfe}
\end{equation}
In the electric limit $a^0, ~a^i, ~a_0, ~a_i$ will transform under Galilean boost as
\begin{eqnarray}
    a'^0 = a^0, \,\,\, a'^i = a^i - u^i a^0, \,\,\, a'_0 = a_0 + u^i a_i, \,\,\, a'_i =a_i \label{gbt1}
\end{eqnarray}
As pointed out in \cite{Bhattacharya}, the electric and magnetic fields transform in the electric limit
\begin{equation}
    e'^i = e^i, \,\,\, ~~b'^i = b^i - \Big(\vec v \times \vec e \Big)^i, \,\,\, ~~e'_i = e_i + \Big( \vec v \times \vec b \Big)_i, \,\,\, ~~b'_i = b_i \label{ebt}
\end{equation}
Using \ref{ebt} it is evident that the Maxwell part of the lagrangian (\ref{lfe}) remains invariant
\begin{equation}
    \delta \mathcal{L}_{Maxwell} = 0 
\end{equation}

We will show here that remaining terms are also boost invariant. To do so we write
\begin{equation}
    \mathcal{L}_{Mass} =  \frac{k^2}{2} \Big(a_0 a^0 + a_i a^i  \Big) \label{lp}
\end{equation}
Now taking the variation of \ref{lp} and using \ref{gbt1} we have
\begin{equation}
 \delta \mathcal{L}_{Mass} =  \frac{k^2}{2} \Big(\delta a_0 a^0 + a_0 \delta a^0 + \delta a_i a^i + a_i \delta a^i  \Big)  = \frac{k^2}{2} \Big( u^i a_i a^0 - u^i a^0 a_i \Big) = 0
\end{equation}
In the magnetic limit $a^0, ~a^i, ~a_0, ~a_i$ will transform under Galilean boost as
\begin{eqnarray}
    a'^0 = a^0 + u_j a^j, \,\,\, a'^i = a^i, \,\,\, a'_0 = a_0 , \,\,\, a'_i =a_i - u_i a_0 \label{gbt2}
\end{eqnarray}
And using \ref{gbt2} we can write the variation of \ref{lp} as
\begin{equation}
   \delta \mathcal{L}_{Mass} = \frac{k^2}{2} \Big(a_0 u_j a^j - a^i u_i a^0  \Big) =0
\end{equation}
Thus the complete lagrangian \ref{lfe} is boost invariant. 
\section{Noether Currents} \label{sec5}
\noindent Contrary to  Maxwell theory there is no conserved current that directly follows from the Noether procedure since there is no gauge invariance. However it is possible to redefine the current so that it is conserved. This happens because although the Lagrangian is not invariant, the action is on-shell invariant. To see this consider the variation,
\begin{equation}
    \delta A_\mu = \6_\mu \alpha
\end{equation}
which modifies the Proca lagrangian to 
\begin{eqnarray}
    \delta \mathcal{L}_{\rm Proca} = k^2 A_\mu \6^\mu \alpha = k^2 \6^\mu \Big( A_\mu \alpha \Big) - k^2 \6^\mu A_\mu \alpha 
\end{eqnarray}
Hence action is on-shell invariant
\begin{equation}
    \delta S = \int \delta \mathcal{L}_{\rm Proca} = 0 
\end{equation}
This indicates the possibility to redefine the current that follows from the Noether procedure such that it is conserved. The usual Noether current is,
\begin{equation}
    J^\mu = \frac{\6 \mathcal{L}}{\6 (\6_\mu A_\nu)} \delta A_\nu = -F^{\mu \nu} \6_\nu \alpha
\end{equation}
Redefining the current as ,
\begin{equation}
    J'^\mu = J^\mu - k^2 A^\mu \alpha
\end{equation}
it is possible to show its on-shell conservation,
\begin{equation}
    \6_\mu J'^\mu = -\6_\mu F^{\mu \nu} \6_\nu \alpha - k^2 A^\mu \6_\mu \alpha - k^2 (\6_\mu A^\mu) \alpha
\end{equation}
Finally using the equations of motion \ref{peom1} and \ref{peom2} we have 
\begin{equation}
    \6_\mu J'^\mu = 0
\end{equation}
Now we see the implications of these currents in the galilean limits. To do this we will start with the electric limit.\\
\noindent \underline{\bf Electric limit:}\\
The map from $J^\mu \to j^\mu$ or $J_\mu \to j_\mu$ just follow the corresponding map for vectors, $A^\mu \to a^\mu$, $A_\mu \to a_\mu$,
\begin{eqnarray}
    J^0 \to c j^0, \,\,\, J^i \to j^i, \,\,\, J_0 \to \frac{j_0}{c}, \,\,\, J_i \to j_i
\end{eqnarray}
Using the above relation we can write the Noether currents in Galilean limit as 
\begin{equation}
 j^0 = \6^i a^0 \6_i \alpha, \,\,\,\,\,\,    j^i = -\6^i a^0 \6_t \alpha - f^{ij} \6_j \alpha 
\end{equation}

Now we define the modified currents $J'^\mu$ which in the non relativistic limit gives rise to,
\begin{eqnarray}
  j'^0 = j^0 - k^2 a^0 \alpha, \,\,\,\,\, j'^i = j^i - k^2 a^i \alpha  
\end{eqnarray}
The conservation equation reduces to
\begin{equation}
    \6_{\mu} J'^{\mu} = 0 \xrightarrow[\text{$c \to \infty$}]{\text{electric limit}} \6_t j'^0 + \6_i j'^i = 0
\end{equation}
To see current conservation in the galilean limit we compute 
\begin{eqnarray}
    \6_t j'^0 + \6_i j'^i = \6_t j^0 + \6_i j^i - \6_t (k^2 a^0 \alpha) - \6_i (k^2 a^i \alpha) = 0
\end{eqnarray}

\noindent In the above equation we have used eqns \ref{ee1}, \ref{ee2} and \ref{ae2}.

\noindent The covariant currents are derived by adopting a similar approach. From the relativistic expression using the mapping relations given in table \ref{T1},
\begin{equation}
j_0 = -\Big(\6_t a_i - \6_i a_0  \Big) \6^i \alpha, \,\,\,\,\,\,    j_i = -f_{ij} \6^j \alpha  
\end{equation}

\noindent The modified currents are given by, 
\begin{eqnarray}
  j'_0 = j_0 - k^2 a_0 \alpha, \,\,\,\,\, j'_i = j_i - k^2 a_i \alpha  
\end{eqnarray}

\noindent The conservation equation reduces to 
\begin{eqnarray}
    \6^\mu J'_\mu = 0 \xrightarrow[\text{$c \to \infty$}]{\text{electric limit}}  \6^i j'_i = 0
\end{eqnarray}
Here 
\begin{eqnarray}
    \6^i j'_i =  \6^i j^i - k^2 \6^ia_i \alpha -k^2 a_i \6^i \alpha = 0
\end{eqnarray}
\noindent In the above equation we have used eqns \ref{ee4} and \ref{ae1}. \\
\noindent \underline{\bf Magnetic limit:}\\
\noindent In this limit,
\begin{eqnarray}
    J^0 \to -\frac{j^0}{c}, \,\,\, J^i \to j^i, \,\,\, J_0 \to -c j_0, \,\,\, J_i \to j_i
\end{eqnarray}
The contravariant components of the currents are
\begin{eqnarray}
    j^0 = \Big( -\6_t a^i + \6^i a^0 \Big) \6_i \alpha, \,\,\, j^i = -f^{ij} \6_j \alpha 
\label{currents1}
\end{eqnarray}
Conservation of the modified currents imply
\begin{eqnarray}
    \6_i j'^i = \6_i j^i - k^2 (\6_i a^i) \alpha - k^2 a^i \6_i \alpha = 0
\end{eqnarray}
where we have exploited eqns \ref{currents1},  \ref{em2} and \ref{am2}.

\section{Stuckelberg embedded version of Proca model} \label{sec6}
The theory is described by the Lagrangian 
\begin{equation}
   \mathcal{L}_{\rm SProca} = -\frac{1}{4}~F_{\mu \nu}~F^{\mu \nu} + \frac{k^2}{2} \Big(A_\mu + \6_\mu \Phi \Big) \Big(A^\mu + \6^\mu \Phi \Big) \label{sl1}
\end{equation}
where  $\Phi$ is a scalar field. We can now see that the Lagrangian is gauge invariant under following transformations,
\begin{equation}
    A_\mu \to A_\mu + \6_\mu \alpha, \,\,\,\,\, \Phi \to \Phi - \alpha
\end{equation}
Now varying the action wrt $A_\nu$ and $\Phi$ we have the following equations of motion, 
\begin{equation}
    \6_\mu F^{\mu \nu} + k^2 \Big( A^\nu + \6^\nu \Phi \Big) = 0, ~~~~~~~~~~~~ \6_\mu \Big( A^\mu + \6^\mu \Phi \Big) = 0 \label{seq1}
\end{equation}

We see that the two equations are self consistent. 
Since our aim is to construct Stuckelberg embedded Galilean Proca model where covariant and contravariant sectors are treated distinctly, it is necessary to write a more general version of \ref{sl1},
\begin{equation}
   \mathcal{L}_{\rm SProca} = -\frac{1}{4}~F_{\mu \nu}~F^{\mu \nu} + \frac{k^2}{2} \Big(A_\mu + \6_\mu \Phi \Big) \Big(A^\mu + \6^\mu \Psi \Big) \label{sl2}
\end{equation}
which bears an extended gauge invariance
\begin{eqnarray}
    A_\mu \to A_\mu + \6_\mu \alpha, \,\,\,\,~~~~~ \Phi \to \Phi - \alpha \label{G1} \\
    A^\mu \to A^\mu + \6^\mu \beta, \,\,\, \, ~~~~~\Psi \to \Psi - \beta \label{G2}
\end{eqnarray}
The equations of motion are obtained by varying $A_\nu, ~A^\nu, ~\Phi ~\& ~\Psi$
\begin{eqnarray}
    \6_\mu F^{\mu \nu} + k^2 \Big(A^\nu + \6^\nu \Psi   \Big) = 0, \\
     \6^\mu F_{\mu \nu} + k^2 \Big(A_\nu + \6_\nu \Phi   \Big) = 0, \\
     \6_\mu \Big(A^\mu + \6^\mu \Psi   \Big) = 0, \\
     \6^\mu \Big( A_\mu + \6_\mu \Phi \Big) =0
\end{eqnarray}
We will now consider different limits, beginning with the electric case. \\
\noindent \underline{\bf Electric limit:}\\
\noindent In this limit we know the fields scale as 
\begin{eqnarray}
    A^0 \to c a^0, \,\, A^i \to a^i, \,\, A_0 \to \frac{a_0}{c}, \,\, A_i \to a_i, \,\, \Phi \to \phi, \,\, \Psi \to \psi
\end{eqnarray}

\noindent The Lagrangian \ref{sl2} imbibes the form,
\begin{equation}
    \mathcal{L}_{\rm SPe} = \frac{1}{2} \6^i a^0 \Big( \partial_t a_i - \partial_i a_0  \big)  - \frac{1}{4} f_{ij} f^{ij} + \frac{k^2}{2} \Big(a_0 a^0 + a_i a^i  \Big) + \frac{k^2}{2} a_i \6^i \psi + \frac{k^2}{2} \Big( a^0 \6_t \phi + a^i \6_i \phi  \Big) + \frac{k^2}{2} \6_i \phi \6^i \psi \label{sel1}
\end{equation}
\noindent Varying with respect to $a_0,~a_j,~a^0,~a^j, ~\phi ~\& ~\psi$ we get the corresponding equations of motion
\begin{eqnarray}
\6_i \6^i a^0 + k^2 a^0 = 0 
\label{see1}
\end{eqnarray}
\begin{eqnarray}
\6_t \6^j a^0 + \6_i \6^j a^i - \6_i \6^i a^j - k^2 \Big( a^j + \6^j \psi \Big) = 0
\label{see2}
\end{eqnarray}
\begin{eqnarray}
\6^i \6_t a_i - \6^i \6_i a_0 - k^2 \Big( a_0 + \6_t \phi \Big) = 0
\label{see3}
\end{eqnarray}
\begin{eqnarray}
\6^i \6_i a_j - \6^i \6_j a_i + k^2 \Big( a_j + \6_j \phi \Big) = 0 
\label{see4}
\end{eqnarray}
\begin{eqnarray}
    \6_t a^0 + \6_i \Big(a^i + \6^i \psi  \Big) = 0
    \label{see5}
\end{eqnarray}
\begin{eqnarray}
    \6^i \Big(a_i + \6_i \phi \Big) = 0
    \label{see6}
\end{eqnarray}
We now derive these equations directly from the equations of motion. The relativistic equations are given by,
\begin{equation}
    \6_{\mu} F^{\mu \nu} + k^2 \Big( A^\nu + \6^\nu \Psi \Big) = 0, ~~~~~~~~~\6_\mu \Big(A^\mu + \6^\mu \Psi   \Big) = 0
\end{equation}
which can be written as
\begin{eqnarray}
\6_i F^{i0} + k^2 \Big( A^0 + \6^0 \Psi \Big)  =0 \label{se1}\\
\6_0 F^{0j} + \6_i F^{ij} + k^2 \Big( A^j + \6^j \Psi \Big) = 0 \label{se2}\\
\6_0 \Big( A^0 + \6^0 \Psi \Big) + \6_i \Big( A^i + \6^i \Psi \Big) = 0 \label{se5}
\end{eqnarray}
Using table \ref{T1} and taking appropriate limits, we reproduce \ref{see1} from \ref{se1}, \ref{see2} from \ref{se2} and \ref{see5} from \ref{se5}.\\
\indent To get the remaining pair of equations we have to interpret  eqns \ref{se1} and \ref{se2} so that the variables appear in a covariant form,
\begin{eqnarray}
\6^i F_{i0} + k^2 \Big( A_0 + \6_0 \Phi \Big) =0 \label{se3}\\
\6^0 F_{0j} + \6^i F_{ij} +k^2 \Big( A_j + \6_j \Phi \Big) = 0 \label{se4}\\
\6^0 \Big( A_0 + \6_0 \Phi \Big) + \6^i \Big( A_i + \6_i \Phi) =0 \label{se6}
\end{eqnarray}
Following similar steps \ref{se3} reproduces \ref{see3} while \ref{se4} yields \ref{see4} and \ref{se6} yields \ref{see6} in the Galilean limit.\\
\noindent \underline{\bf Magnetic limit:}\\
\noindent In this limit we know the fields scale as 
\begin{eqnarray}
    A^0 \to -\frac{a^0}{c}, \,\, A^i \to a^i, \,\, A_0 \to -c a_0, \,\, A_i \to a_i, \,\, \Phi \to \phi, \,\, \Psi \to \psi
\end{eqnarray}
Now the Lagrangian looks like 
\begin{equation}
    \mathcal{L}_{\rm SPm} = \frac{1}{2} \6_i a_0 \Big( \partial_t a^i - \partial^i a^0  \big)  - \frac{1}{4} f_{ij} f^{ij} + \frac{k^2}{2} \Big(a_0 a^0 + a_i a^i  \Big) + \frac{k^2}{2} a^i \6_i \phi + \frac{k^2}{2} \Big( a_0 \6_t \psi + a_i \6^i \psi  \Big) + \frac{k^2}{2} \6_i \phi \6^i \psi \label{sml1}
\end{equation}
Varying \ref{sml1} wrt $a_0,~a_j,~a^0,~a^j, ~\phi, ~\psi$ we get,
\begin{eqnarray}
   \partial_i ~\partial_t~ a^i - \partial^i ~\partial_i ~a^0 - k^2 \Big( a^0 + \6_t \psi \Big) = 0 
  \label{sem1}
\end{eqnarray}
\begin{eqnarray}
   \6_i \6^i a^j - \6_i \6^j a^i + k^2 \Big( a^j + \6^j \psi \Big) = 0
  \label{sem2}
  \end{eqnarray}
\begin{eqnarray}
  \6^i \6_i a_0 + k^2 a_0 = 0
  \label{sem3}
\end{eqnarray}
\begin{eqnarray}
 \6_t \6_j a_0 + \6^i \6_j a_i - \6^i \6_i a_j - k^2 \Big( a_j + \6_j \phi \Big) = 0 \label{sem4}
\end{eqnarray}
\begin{eqnarray}
    \6_i \Big(a^i + \6^i \psi \Big) = 0
    \label{sem5}
\end{eqnarray}
\begin{equation}
    \6_t a_0 + \6^i \Big( a_i + \6_i \phi \Big) = 0 \label{sem6}
\end{equation}
Here also we can show that the above equations agree with those derived directly from relativistic Maxwell equations by taking the magnetic limit. Using table \ref{T1} and taking appropriate limits, we reproduce \ref{sem1} from \ref{se1} and \ref{sem2} from \ref{se2}.\\
\noindent To get the remaining pair of equations we have to start from the covariant versions (\ref{se3}, \ref{se4}). 

\begin{table}
\caption{Field equations}\label{T3}
\begin{center}
\begin{tabular}{|c|c|} \hline 
${\rm Electric ~~limit}$ & ${\rm Magnetic ~~limit}$ \\ \hline
  $\6^i \6_t a_i - \6^i \6_i a_0 -k^2 \Big(a_0 + \6_t \phi \Big) = 0$ & $\6^i \6_i a_0 +k^2 a_0 = 0$ \\ \hline
$\6^i \6_i a_j - \6^i \6_j a_i + k^2 \Big(a_j + \6_j \phi \Big) = 0  $ & $\6_t \6_j a_0 + \6^i \6_j a_i - \6^i \6_i a_j - k^2 \Big( a_j + \6_j \phi \Big) = 0$ \\ \hline
$ \partial^i \partial_i a^0 +k^2 a^0 = 0 $ & $\partial_i ~\partial_t~ a^i - \partial^i ~\partial_i ~a^0 - k^2 \Big( a^0 + \6_t \psi \Big)  = 0 $ \\ \hline
$\6_t \6^j a^0 + \6_i \6^j a^i - \6_i \6^i a^j - k^2 \Big( a^j + \6^j \psi \Big) = 0$ & 
$ \6_i \6^i a^j - \6_i \6^j a^i + k^2 \Big( a^j + \6^j \psi \Big)  = 0 $ \\ \hline
$ \6_t a^0 + \6_i \Big(a^i + \6^i \psi  \Big) = 0$ &  $\6_i \Big(a^i + \6^i \psi \Big) = 0$ \\ \hline
$ \6^i \Big(a_i + \6_i \phi \Big) = 0$ & $\6_t a_0 + \6^i \Big( a_i + \6_i \phi \Big) = 0$\\ \hline
\end{tabular}
\label{T3}
\end{center}
\end{table}
From table \ref{T3} we can observe that
\begin{equation*}
{\rm   Contravariant \longleftrightarrow Covariant} \implies {\rm Electric \longleftrightarrow Magnetic} ~~\& ~~\phi \longleftrightarrow \psi
\end{equation*}
\subsection{Gauge invariance}
We have already shown that the relativistic lagrangian is gauge invariant. Here we show that the Galilean counterpart is also gauge invariant. Electric and magnetic limits are treated separately.\\
\noindent \underline{\bf Electric limit:}\\
\noindent In this limit the relations given in \ref{G1} and \ref{G2} will give rise
\begin{equation}
    a^0 \to a^0, \,\,\, a^i \to a^i + \6^i \beta, \,\,\, a_0 \to a_0 + \6_t \alpha, \,\,\, a_i \to a_i + \6_i \alpha \label{g1}
\end{equation}
Using eqn \ref{g1} we can show that
\begin{equation}
    \mathcal{L}_{\rm SPe} \to \mathcal{L}_{\rm SPe}
\end{equation}
\noindent \underline{\bf Magnetic limit:}\\
\noindent In this limit the relations given in \ref{G1} and \ref{G2} will give rise
\begin{equation}
    a^0 \to a^0 + \6_t \beta, \,\,\, a^i \to a^i + \6^i \beta, \,\,\, a_0 \to a_0, \,\,\, a_i \to a_i + \6_i \alpha \label{g2}
\end{equation}
Using eqn \ref{g2} we can show that
\begin{equation}
    \mathcal{L}_{\rm SPm} \to \mathcal{L}_{\rm SPm}
\end{equation}
\subsection{Noether currents}
\noindent For the relativistic Stuckelberg embedded Proca model the Noether current is defined as
\begin{equation}
    J^{\mu} = -F^{\mu \nu} \6_{\nu} \alpha - k^2 \alpha \Big( A^\mu + \6^\mu \Psi \Big) \label{J1}
\end{equation}
One can clearly see that this current is conserved
\begin{equation}
    \6_\mu J^\mu = 0 \label{con1}
\end{equation}
we will now discuss the Galilean limit \\
\noindent {\underline {\bf Electric limit:}}\\
Relativistic conservation equation in the non-relativistic (electric) limit, reduces to,
\begin{equation}
     \6_\mu J^\mu = 0 \xrightarrow[\text{$c \to \infty$}]{\text{electric limit}} \6_t j^0 + \6_i j^i = 0
\end{equation}
where using eqn \ref{J1} and the mapping relations given in table \ref{T1} we can write down the current components as
\begin{equation}
    j^0 = \6^i a^0 \6_i \alpha - k^2 \alpha a^0, \,\,\,\, ~~~~ j^i = \6^i a^0 \6_t \alpha - f^{ij} \6_j \alpha - k^2 \alpha \Big( a^i + \6^i \psi \Big) \label{j1}
\end{equation}
\noindent This conservation follows by using eqns \ref{see1}, \ref{see2}, \ref{see5} and \ref{j1}.\\
\indent Following the same steps we can show current conseravation for covariant components. \\
\noindent {\underline {\bf Magnetic limit:}}\\
\noindent Relativistic conservation equation in the magnetic limit, reduces to,
\begin{equation}
     \6_\mu J^\mu = 0 \xrightarrow[\text{$c \to \infty$}]{\text{magnetic limit}} \6_i j^i = 0
\end{equation}
where using eqn \ref{J1} and the mapping relations given in table \ref{T1} we can write down the current components as
\begin{equation}
    j^0 = -\Big(\6_t a^i - \6^i a^0  \Big) - k^2 \alpha \Big( a^0 + \6_t \psi \Big), \,\,\,\, ~~~~~~~~~j^i = -f^{ij} \6_j \alpha - k^2 \alpha \Big( a^i + \6^i \psi \Big) \label{j2}
\end{equation}
This conservation follows by using eqns \ref{sem1}, \ref{sem2}, \ref{sem5} and \ref{j2}.

\section{Conclusions} \label{sec7}
\noindent Let us now summarise the new significant findings of the paper, comparing with existing results found in the literature. 
\begin{itemize}
    \item An unambiguous construction of the action principle for non-relativistic (NR) Proca model, for both electric and magnetic limits, was given. The non-relativistic analogue of massive spin-1 particle has been discussed in \cite{Leblond2} by using a group theory point of view and a short discussion has been presented in \cite{Santos}. None of these is comprehensive and does not include the several features presented here. For instance in \cite{Leblond2} only the magnetic limit is considered while \cite{Santos} presents a mixed lagrangian with both electric and magnetic sectors. Nowhere a systematic construction of an action principle was presented. Here we have deduced this action from the standard relativistic lagrangian adopting the dictionary given in section \ref{sec2}. It is expressed either in terms of potentials or fields. Explicit demonstration of boost invariance has been presented.
    \vspace{0.1in}
    \item We have considered contravariant and covariant vectors separately. This helps us to discover certain subtleties which were unobserved so far. For example it is observed from table \ref{T2} that if we replace the covariant components by contravariant ones in the electric limit case we will end up with the magnetic limit case and vice-versa.
    \vspace{0.1in}
    \item A central point is the formulation of a dictionary that translates four vectors in the relativistic theory to their corresponding vectors in the non-relativistic theory. Also, we know gauge symmetries play a pivotal role in the understanding of gauge theories. Since there is no gauge symmetry present in the Proca theory it is interesting to observe its consequence in the Galilean invariant theory which has been derived by using the dictionary presented here. We also discuss in section \ref{sec6} how to restore the gauge invariance by using the Stuckelberg embedding. We derive its Galielan counterpart by writing down the lagrangian for both the limits. Since covariant and contravariant components are treated separately, two scalars have to be introduced in the Galilean theory to attain gauge invariance. We show that the Galilean lagrangians for both the limits are gauge invariant. Another interesting thing to observe is the change of contravariant indices by covariant ones induces not only the electric limit to magnetic limit but the Stuckelberg scalars $\phi$ and $\psi$ also get interchanged. 
\end{itemize}
\noindent \underline{\bf Future prospects:}
\begin{itemize}
    \item There have been some observational aspects of Proca models in the context of  gravitational waves \cite{Radu1} and black hole shadows \cite{Radu2}. We believe a further systematic analysis of the non-relativistic Proca theory can shed some important light in this direction.
    \vspace{0.1in}
    \item  Recently its has been observed in \cite{derham} that a certain type of  non-linear Proca field theory which is sometimes called Proca-Nuevo model plays a crucial role in explaining the emergence of half degrees of freedom in Lorentz and parity invariant field theories. It will be interesting to look at the Galilean invariant counterpart and study its further implications. 
    \vspace{0.1in}
    \item Usual Proca model can be generalised by considering derivative
self-interactions with only three propagating degrees of freedom \cite{Heisenberg}. This is sometimes called generalised Proca model. It will be useful to study its Galilean invariant counterpart. The generalised Proca model can be extended to curved spacetime which gives rise to Horndeski Proca model. We like to study the non-relativistic analogue (in a curved background) by adopting the approach of \cite{Rabin1, Rabin2}. 
\vspace{0.1in}
 \item Generalised Proca theories of gravity represent an interesting class of vector-tensor theories which can be extended to modified theories of gravity by including torsion \cite{Said}. Considering the recent interests and applications of non-Lorentzian physics in different contexts, we believe the study of the non-relativistic limit will illuminate various unknown aspects of generalised Proca model as well as  modified gravity theory.
\vspace{0.1in}
\item Finally we want to look at the Carrollian limit \cite{Duval} of the Proca model. There have been some interesting studies of Carroll symmetries in the context of gravitational waves \cite{Duval2} and in the context of Hall effect \cite{Horvathy}. We would like to study the Carrollian limit for these issues, particularly the gravitational waves aspect, in the context of Proca model.  
\end{itemize}
We expect we can address these issues in the near future.
\section{Acknowledgements}
\noindent The authors (RB and SB) acknowledge the support from a DAE Raja Ramanna Fellowship (grant no: $1003/(6)/2021/RRF\\/R\&D-II/4031$, dated: $20/03/2021$).


\begin{thebibliography}{99}
\bibitem{Taylor} M. Taylor, Non-relativistic holography, arXiv:hep-th/0812.0530. 
\bibitem{andreev1}  O. Andreev, M. Haack and S. Hofmann, Phys. Rev. D 89, 064012
(2014) doi:10.1103/PhysRevD.89.064012 [arXiv:1309.7231 [hep-th]].
\bibitem{andreev2}  O. Andreev, Phys. Rev. D 91, no. 2, 024035 (2015)
doi:10.1103/PhysRevD.91.024035 [arXiv:1408.7031 [hep-th]].
\bibitem{jensen}  K. Jensen and A. Karch, JHEP 1504, 155 (2015)
doi:10.1007/JHEP04(2015)155 [arXiv:1412.2738 [hep-th]].
\bibitem{rb1} R. Banerjee and P. Mukherjee, Phys. Rev. D 93, no. 8, 085020 (2016)
doi:10.1103/PhysRevD.93.085020 [arXiv:1509.05622 [gr-qc]].
\bibitem{rb2}  R. Banerjee, S. Gangopadhyay and P. Mukherjee, Int. J. Mod. Phys.
A 32, no. 19n20, 1750115 (2017) doi:10.1142/S0217751X17501159
[arXiv:1604.08711 [hep-th]].
\bibitem {Pal} B. Grinstein and S. Pal, Phys. Rev. D 97, no. 12, 125006 (2018).
\bibitem{Jain} A. Jain, Phys. Rev. D 93, no. 6, 065007 (2016).
\bibitem{rb3} R. Banerjee and P. Mukherjee, Subtleties of nonrelativistic reduction and applications, Nucl. Phys. B {\bf 938}, 1 (2019), [arXiv: 1801.08373].
\bibitem{Morand} K. Morand, Embedding Galilean and Carrollian geometries I. Gravitational waves, Journal of Mathematical Physics 61, 082502 (2020), [arXiv:1811.12681 [hep-th]].
\bibitem{Leblond} M. L. Bellac and J.-M. Levy-Leblond, Galilean Electromagnetism, Nuovo Cimento. 14B (1973) .
\bibitem{Sengupta}  N. D. Sengupta, On an Analogue of the Galilei Group,  Nuovo Cim. 54 (1966) 512, DOI: 10.1007/BF02740871.
\bibitem{Germain} G. Rousseaux, Forty years of Galilean Electromagnetism (1973–2013), Eur. Phys. J. Plus (2013) 128: 81.
\bibitem{Duval} C. Duval, G. W. Gibbons, P. A. Horvathy and P. M. Zhang, Carroll versus Newton and Galilei: two dual non-Einsteinian concepts of time, Class. Quant. Grav. 31 (2014) 085016
[1402.0657].
\bibitem{Mehra1} A. Bagchi, R. Basu and A. Mehra, Galilean Conformal Electrodynamics, JHEP 11 (2014) 061 [1408.0810].

\bibitem{Bhattacharya} R. Banerjee and S. Bhattacharya, New formulation of Galilean relativistic Maxwell theory, Phys. Rev. D. 107, 105022 (2023), [arXiv: hep-th/2211.12023].
\bibitem{Proca} A. Proca, J. Phys. Radium 7, 347 (1936).
\bibitem{Luo} L.C. Tu, J. Luo, G.T. Gillies, The mass of the photon, Rep. Prog. Phys. 68, 77 (2005).
\bibitem{Nieto} A.S. Goldhaber, M.M. Nieto, Photon and graviton mass limits, Rev. Mod. Phys. 82, 939 (2010).
\bibitem{Sampaio} C. Herdeiro, M.O.P. Sampaio, M. Wang, Hawking radiation for a Proca field in $D$ dimensions, Phys. Rev. D 85, 024005 (2012).
\bibitem{Dvali} G. Dvali, M. Papucci, M.D. Schwartz, Infrared Lorentz Violation and Slowly Instantaneous Electricity, Phys. Rev. Lett. 94, 191602 (2005).
\bibitem{Tomaschitz}  R. Tomaschitz, Tachyonic spectral fits of $\gamma$-ray bursts, Europhys. Lett. 89, 39002 (2010).
\bibitem{Radu1}  N. Sanchis-Gual, J. C. Bustillo, C. Herdeiro, E. Radu, J. A. Font, S. H. W. Leong, A. Torres-Forné, Impact of the wave-like nature of Proca stars on their gravitational-wave emission, Phys. Rev. D 106, 124011 (2022), [arXiv: 2208.11717].
\bibitem{Radu2}  I. Sengo, P. V. P. Cunha, C. A. R. Herdeiro, E. Radu, Kerr black holes with synchronised Proca hair: lensing, shadows and EHT constraints, JCAP 01 (2023) 047.
\bibitem{Demir}  D. Demir and B. Pulice,  Geometric Proca with Matter in Metric-Palatini Gravity,
 Eur. Phys. J. C 82, 996 (2022).
 \bibitem{Heisenberg} L. Heisenberg, Generalization of the Proca Action, JCAP 1405 (2014) 015 [arXiv: 1402.7026].
 \bibitem{Said}  G. P. Nicosia, J. L. Said and V. Gakis, Generalised Proca Theories in Teleparallel Gravity, EPJP, volume 136, Article number: 191 (2021), https://
doi.org/10.1140/epjp/s13360-021-01133-4.
 \bibitem{Pichet}  J. Sanongkhun and P. Vanichchapongjaroen, On constrained analysis and diffeomorphism invariance of generalised Proca theories, General Relativity and Gravitation volume 52, Article number: 26 (2020), https://doi.org/10.
1007/s10714-020-02678-y.
 \bibitem{derham} C. de Rham, S. Garcia-Saenz, L. Heisenberg, V. Pozsgay and X. Wang,  To Half–Be or Not To Be?, JHEP 06, 2023, 88 (2023), [arXiv: 2303.05354]. 
\bibitem{Santos} E. S. Santos, M. de Montigny, F. C. Khanna and A. E. Santana, Galilean covariant Lagrangian models, J. Phys. A 37 (2004) 9771.
\bibitem{Weinberg} S. Weinberg, The Quantum Theory of Fields, Vol-I, Page-62, Cambridge University Press, (1995). doi:10.1017/CBO9781139644167.
\bibitem{Leblond2} Lévy-Leblond J. M. 1967 Nonrelativistic particles and wave equations Commun. Math. Phys. 6 286–311.

\bibitem{Rabin1} R. Banerjee, A. Mitra and P. Mukherjee, A new formulation of non-relativistic diffeomorphism invariance,  Phys.Lett.B 737 (2014) 369-373, [arXiv: 1404.4491 [gr-qc]].
\bibitem{Rabin2} R. Banerjee, A. Mitra and P. Mukherjee, Localization of the Galilean symmetry and dynamical realization of Newton-Cartan geometry, Class.Quant.Grav. 32 (2015) 4, 045010, [arXiv: 1407.3617 [hep-th]].
\bibitem{Duval2} C. Duval, G. W. Gibbons, P. A. Horvathy, P.-M. Zhang, Carroll symmetry of plane gravitational waves, Class. Quantum Grav. 34 175003 (2017), [arXiv:gr-qc/1702.08284].
\bibitem{Horvathy} L.Marsot, P.-M. Zhang, M. Chernodub, P.A. Horvathy, Hall motions in Carroll dynamics, Phys. Rept. 1028 (2023), [arXiv:hep-th/2212.02360].



\end{thebibliography}
\end{document}